\documentclass[10pt,preprint,aps,prc,showpacs]{revtex4-1}
\usepackage{graphicx,epstopdf,multirow,amsmath,rotating,enumitem}

\begin{document}

		\author{Deepak Kumar}
		\author{Moumita Maiti\footnote{E-mail: moumifph@iitr.ac.in, moumifph@gmail.com (Reprint author)}}
    	\affiliation{Department of Physics, Indian Institute of Technology Roorkee, Roorkee-247667, Uttarakhand, INDIA}

	\title{ Incomplete fusion analysis of $^{7}$Li-induced reaction on $^{93}$Nb within 3-6.5 MeV/nucleon}
	\date{\today}

		\begin{abstract}
		\begin{description}
\item[Background]
	 It is understood from the recent experimental studies that prompt/resonant breakup, and transfer followed by breakup in the weakly bound $^{6,7}$Li-induced reactions play a significant role in the complete-incomplete fusion (CF-ICF), suppression/enhancement in the fusion cross section around the Coulomb barrier.\  
\item[Purpose]
	Investigation of ICF over CF by measuring cross sections of the populated residues, produced via different channels in the $^{7}$Li-induced reaction on $^{\text{nat}}$Nb target within 3--6.5 MeV/nucleon energy region.  
\item[Method] 
	The $^{7}$Li beam was allowed to hit the self-supporting $^{93}$Nb targets, backed by the aluminium (Al) foil alternately, within 3--6.5 MeV/nucleon energy. Populated residues were identified by offline $\gamma$-ray spectrometry. Measured excitation functions of different channels were compared with different equilibrium (EQ) and preequilibrium (PEQ) models. 

\item[Result] 
The enhancement in cross sections in the proton ($\sim$ 20--30 MeV) and $\alpha$-emitting channels, which may be ascribed to ICF, was observed in the measured energy range when compared to the Hauser-Feshbach and exciton model calculations using EMPIRE, which satisfactorily reproduces the neutron channels, compared to the Weisskopf-Ewing model and Hybrid Monte Carlo calculations.
The increment of the incomplete fusion fraction was observed with rising projectile energy. 
			
\item[Conclusion] 
Contrary to the Alice14, experimental results are well reproduced by the EMPIRE throughout the measured energy range. The signature of ICF over CF indicates that the breakup/transfer processes are involved in the weakly bound $^{7}$Li-induced reaction on $^{93}$Nb slightly above the Coulomb barrier.
     \end{description}
	\end{abstract}
   
  \maketitle

	
	\section{\label{s1}Introduction}
	
A number of theoretical and experimental investigations have been carried out in the recent years to disentangle the breakup mechanism of weakly bound projectiles, predominantly in the $^{6,7}$Li ions consisting of $\alpha$-particle + triton ($t$)/deuteron ($d$) cluster structure with a separation energy of 2.47/1.47 MeV, respectively \cite{pakou17,lei17,luong11,srivastava13,capurro16,pandit16,parkar16}. Dominance of large $\alpha$-yields from the non-elastic breakup (NEBU) compared to the elastic breakup (EBU) has been observed in the $^{6,7}$Li projectiles \cite{lei17}.

Breakup of $^{7}$Li projectile may be possible through the following: (i) excitation of relatively long lived resonant state in the projectile continuum followed by its decay into an $\alpha$-particle + $t$, termed as sequential breakup, (ii) direct breakup without any intermediate resonant state, and (iii) breakup into $^{6}$He + $p$ and $^{5}$He + $d$ configuration and subsequently fusion of $^{6}$He or $^{5}$He as a massive transfer process \cite{luong11,srivastava13}. Moreover, transfer of nucleon followed by breakup has also been reported where $\alpha$ + $\alpha$ or $\alpha$ + $d$ breakup occurred after the proton ($p$) pickup by the projectile or neutron ($n$) transfer from the projectile ($^7$Li).  However, it is also reported that the  $\alpha$ + $\alpha$ events from one proton pickup is less probable compared to the $\alpha$ + $d$ events from one neutron stripping and/or $\alpha$ + $t$ event from the direct breakup of $^{7}$Li in the $^{7}$Li + $^{93}$Nb reactions at energies close to Coulomb barrier \cite{capurro16,pandit16}. Further, Parkar \textit{et al.} \cite{parkar16} concluded that $\alpha$--ICF/$\alpha$--capture is less probable compare to triton($t$)--ICF/$t$--capture for $^{7}$Li induced reactions on $^{209}$Bi and $^{198}$Pt targets.

In order to explore the breakup threshold anomaly, several $^{6,7}$Li induced reactions on the light, medium and heavy mass nuclei has been performed around the Coulomb barrier \cite{ray08,shaikh14,padron02,pietro13,gomes05,rath13,dasgupta02,palshetkar14,raabe06,zagatto17,badran02,dm17,dm16}. In many light mass nuclei such as $^{27}$Al, $^{28}$Si, $^{24}$Mg, $^{64}$Ni etc.,\ slight enhancement, but no suppression in the total fusion (TF) cross section was observed below and above the barrier using $^{6,7}$Li beam \cite{ray08,shaikh14,padron02}, respectively. Similarly, Pietro \textit{et al.} \cite{pietro13} reported the enhancement in the TF cross section below the barrier and no suppression above it in $^{6,7}$Li + $^{64}$Zn, contrary to Gomes \textit{et al.}\ \cite{gomes05}. We have observed a significant enhancement in cross section in the $\alpha$-emitting channels of $^{7}$Li + $^{\text{nat}}$Mo reaction which is attributed to the incomplete fusion (ICF) slightly above the Coulomb barrier \cite{dm17}. A significant enhancement and suppression of the complete fusion (CF) cross section below and above the barrier was also reported comparing with the coupled channel calculations in $^{7}$Li induced reactions on the heavy mass nuclei such as $^{209}$Bi, $^{144,154}$Sm, $^{197}$Au, $^{238}$U etc.\cite{rath13,dasgupta02,palshetkar14,raabe06}.  
Zagatto \textit{et al.}\ reported recently the overall good agreement of experimental data with the inclusion of $^{7}$Li coupling to continuum in $^{7}$Li + $^{120}$Sn \cite{zagatto17}.
Investigation of the CF and ICF from the product residues in $^{7}$Li + $^{56}$Fe reaction was examined by observing inclusive light particle spectra along with recoil range distribution of residual products \cite{badran02}. 

Recently, we have reported the cross sections of the residual radionuclides produced in the $^{7}$Li + $^{93}$Nb reaction in the 20-45 MeV energy range \cite{dm16}. As a part of our continuous endeavor, in this article for the first time, a critical analysis of the proton and $\alpha$-emitting channels of the $^{7}$Li + $^{93}$Nb reaction has been reported that indicates the signature of ICF little above the Coulomb barrier.

		
    \section{\label{s2}Experimental}
The experiment was performed at the BARC-TIFR Pelletron facility, Mumbai, India. The $^{7}$Li beam was allowed to impinge on a stack of self-supporting $^{\text{nat}}$Nb metal foils having uniform thickness of 2.3--3.2 mg/cm$^2$. Each target foils were backed by $\sim$1.5 mg/cm$^2$ thin aluminum (Al) foils which served as recoils catcher and energy damper. After the end of the bombardment (EOB), identification of the radionuclides produced at each irradiated target foils was carried out using $\gamma$-ray spectrometry with the help of an HPGe detector and the GENIE-2K software. The yield of the identified residues was determined from the background subtracted peak area count rate. Yields were calculated for each characteristic $\gamma$-peak of the residues observed in the time resolved $\gamma$-ray spectra taken over a long period of time with intervals to assure the complete decay profile of a residual product and were further utilized for the cross section estimation. Detailed discussion of experimental procedure, yield measurement, cross section calculations and associated uncertainties are described elsewhere in our previous articles \cite{dm16,mm11,maiti11,mm10,mm11chem,mm15,mm13}.  
	
	\section{Model calculation}
	
In order to extract information on the nuclear reaction processes involved in the  production of residues via different reaction channels in $^{7}$Li-induced reaction on $^{93}$Nb target, theoretical model calculations have been executed using the upgraded version of nuclear reaction model codes:  Alice14 \cite{blann96} and EMPIRE3.2 \cite{herman07}. Hybrid Monte Carlo Simulation (HMS) \cite{blann96} and Weisskopf-Ewing model (WE) \cite{we40} were utilized for preequilibrium and equilibrium processes in Alice14. Two different level densities such as Fermi gas model (FG) and Kataria Ramamurthy (KR) were used for cross section estimation. Level density parameter, $k$ = 9 and mean free path parameter, $\lambda$ = 1.5 was used in the calculation. On the other hand, EMPIRE uses Hauser-Feshbach formalism (HF) for compound evaporation and the exciton model (EM) with Iwamoto Harada cluster emission model was selected for precompound processes. For heavy-ion induced reactions, simplified coupled channel calculation (CCFUS code) \cite{dasso87} was implemented for the fusion process. Generalized superfluid model (GSM) and enhanced generalized superfluid model (EGSM) was chosen for the nuclear level density, which include the dependence of rotation and vibrational effects on level density. Ignatyuk energy-dependent level density parameter, which considers the dependence of shell effects on the excitation energy, was used in the GSM and EGSM. More details on the models used in the EMPIRE code is described in the articles \cite{dm16,deepak17}.

	 
	\section{\label{s3}Results}

\begin{figure*}[t]
	\includegraphics[width=120mm]{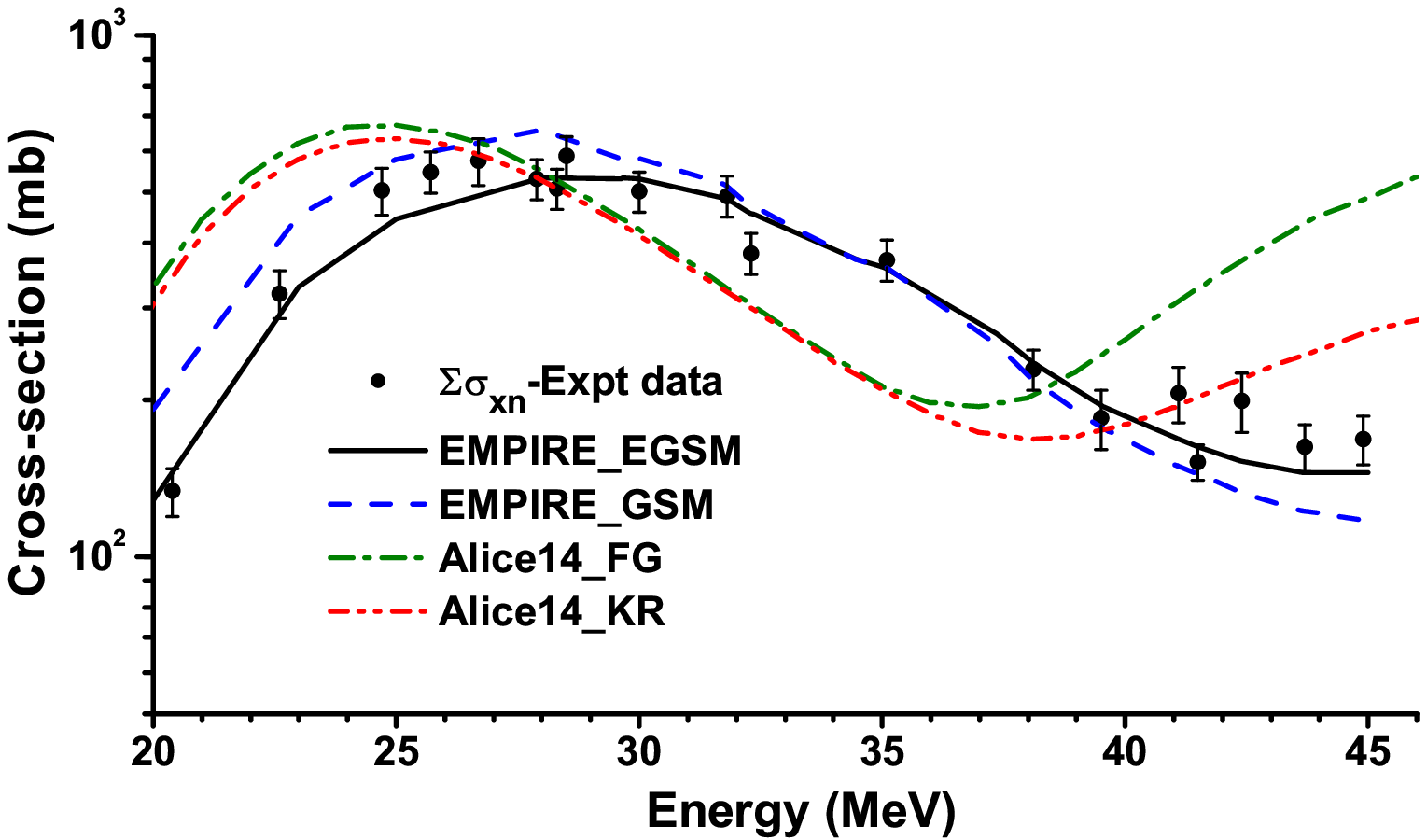}
	\caption{\label{fig1} (Color Online) Comparison of experimental and theoretical excitation functions for the sum of $xn$-channels cross-section.}
\end{figure*}

\begin{figure*}[t]
	\includegraphics[width=120mm]{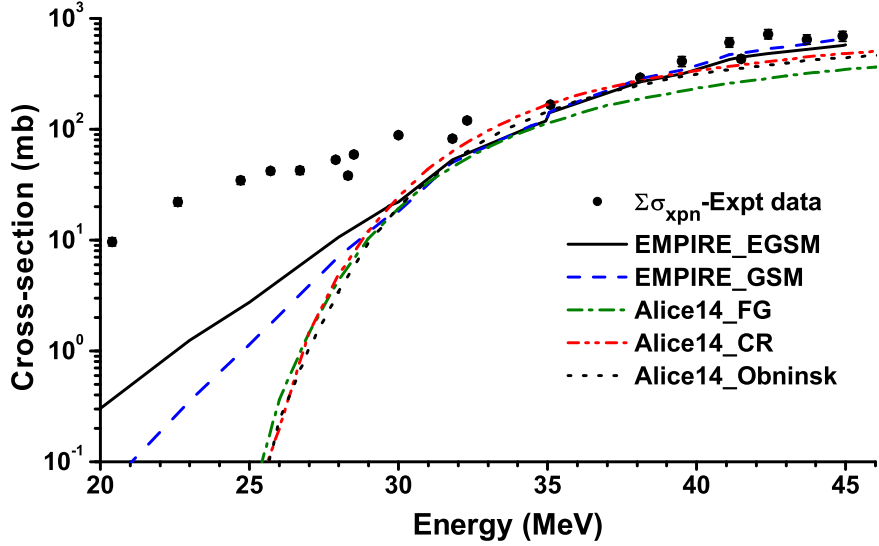}
	\caption{\label{fig2} (Color Online) Same as Fig. \ref{fig1} for the sum of $pxn$-channels cross-section.}
\end{figure*}

\begin{figure*}[t]
	\includegraphics[width=120mm]{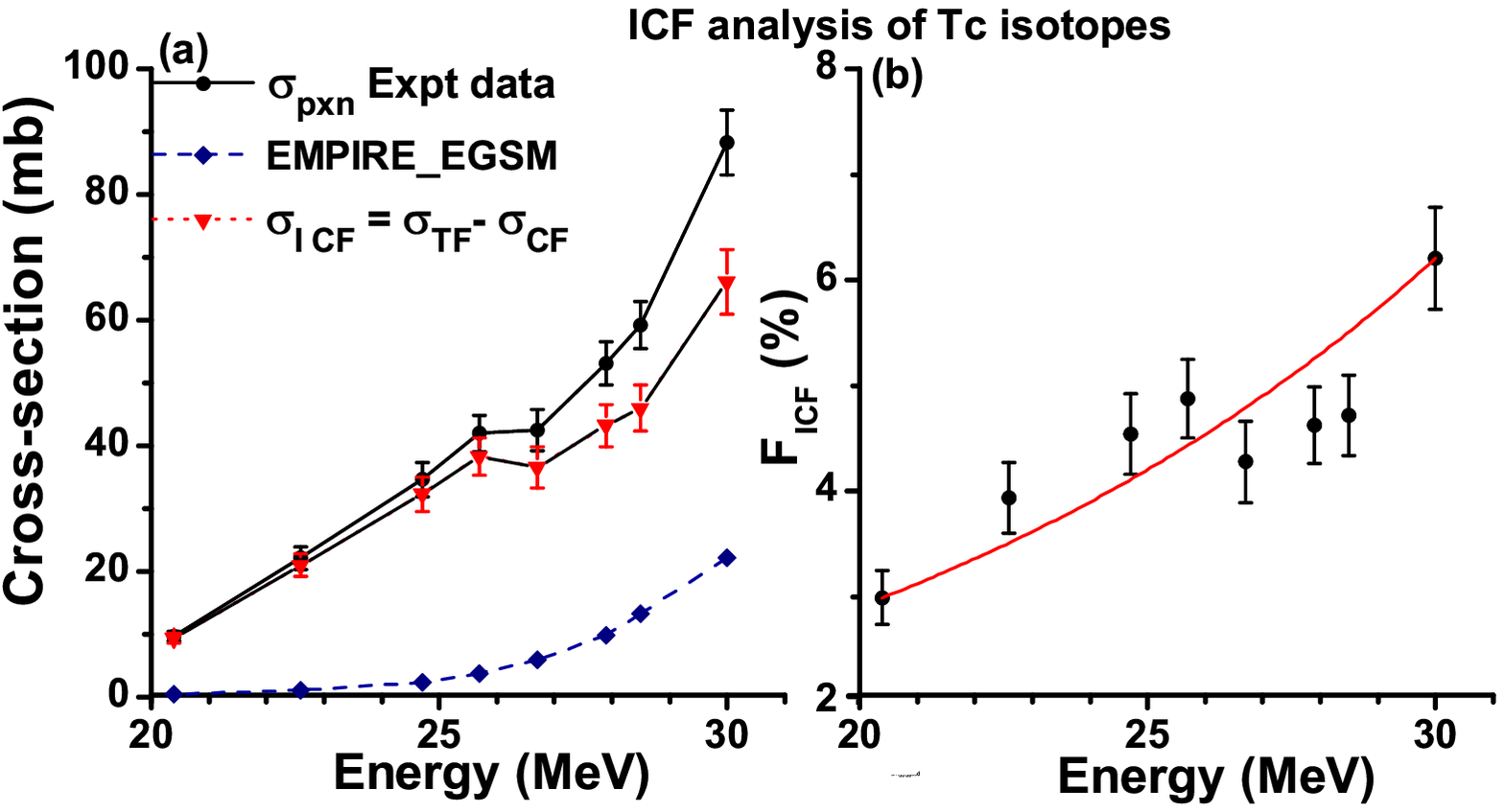}
	\caption{\label{fig3} (Color Online) Variation of complete-incomplete (CF-ICF) fusion cross-section and ICF fraction at various projectile energies. Line fitted through the data in \ref{fig3}(b) is to guide the eye.}
\end{figure*}

\begin{figure*}[t]
	\includegraphics[width=120mm]{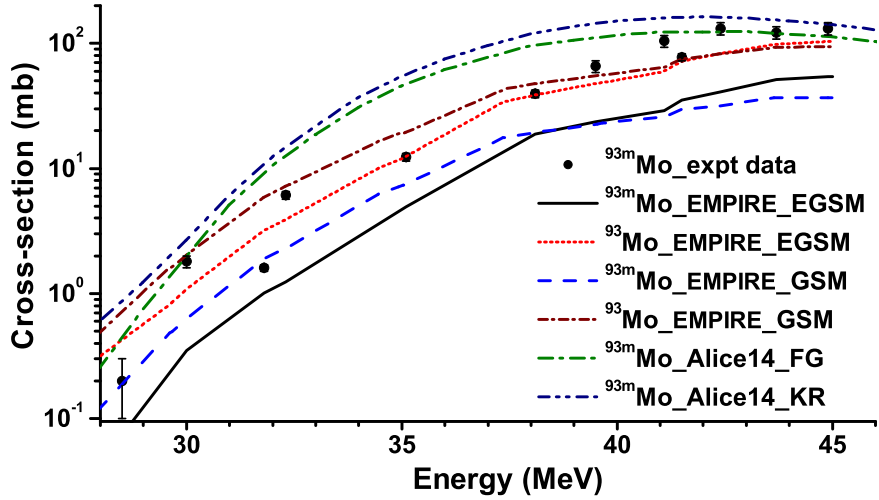}
	\caption{\label{fig4} (Color Online) Same as Fig. \ref{fig1} for the sum of $\alpha 3n$-channel cross-section.}
\end{figure*}

\begin{figure*}[t]
	\includegraphics[width=120mm]{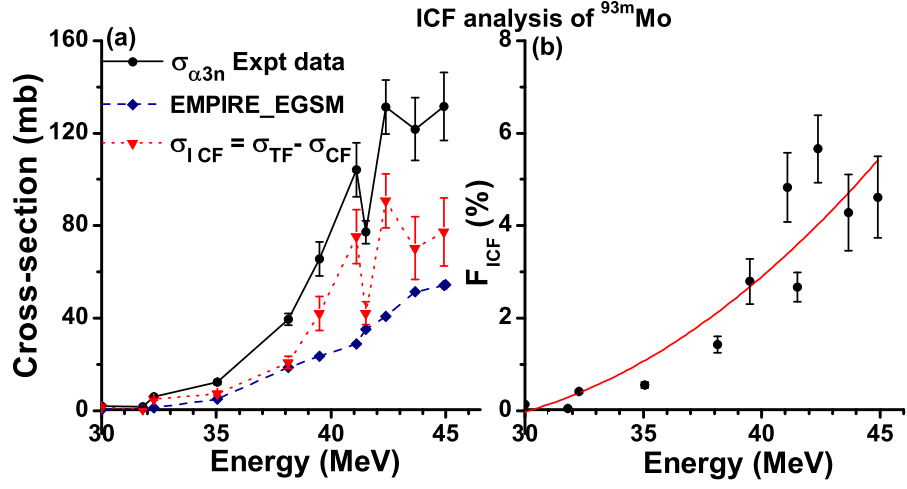}
	\caption{\label{fig5} (Color Online) Variation of complete-incomplete (CF-ICF) cross-section and ICF fraction at various projectile energies. Line fitted through the data in \ref{fig5}(b) is to guide the eye.}
\end{figure*}

It is worthy to note that in our previous article \cite{dm16}, cross sections of all the identified residues: $^{97}$Ru, $^{95}$Ru, $^{96}$Tc, $^{95}$Tc, $^{\text{93m}}$Mo were reported and compared with the theoretical model calculations using PACE4, EMPIRE3.2 and Alice91 in order to extract information about nuclear reaction dynamics such  as PEQ and EQ emissions, etc.
In this report, a systematic analysis of CF-ICF cross section of $\alpha$-particle and proton emitting channels of the $^{7}$Li +$^{93}$Nb reaction has been performed in the $\sim$ 3--6.5 MeV/nucleon energy range. The total cross sections of neutron-, proton-, and $\alpha$-emitting channels are compared with the theoretical model calculations using the upgraded version of nuclear reaction model codes-- EMPIRE3.2 and Alice14, as shown in the Figs. \ref{fig1}, \ref{fig2}, and \ref{fig4}, respectively. Moreover, an analysis of CF--ICF cross section and the incomplete fusion fraction (F$_{ICF}$) at different projectile energies is carried out, as depicted in Figs. \ref{fig3}, \ref{fig5}. 
A detail description of the contributing reactions to populate the residues and their measured cross sections at various projectile energies are reported in the article \cite{dm16}. Measured cross sections are represented by symbols with experimental uncertainties and theoretical results are represented by curves in the figures.

The measured excitation function, the sum of cross sections of the Ru radioisotopes populated via $xn$ channels, are well reproduced by HF + EM calculations with the EGSM level density (Fig \ref{fig1}). However, WE + HMS calculations accomplished using two different level density options-- Fermi gas (FG) and Kataria Ramamurthy (KR) level density follow a different trend relative to the measured data.
On the other hand, in the $pxn$ channels (Fig \ref{fig2}), WE + HMS estimations with FG level density underpredict the measured data throughout the entire energy range; while that with KR level density satisfactorily reproduce them above $\sim$ 32 MeV. The HF + EM calculations reproduce the measured cross section towards the high energy side ($>$32 MeV) and underpredict them in the lower energy region ($<$32 MeV); however, it follows a similar trend of the measured cross section.\ A critical observation reveals that the measured data are well reproduced by EMPIRE with EGSM level density compared to other model calculations except in the lower energy region. This observed enhancement in the measured cross section at the lower energies may be due to the ICF of direct breakup component $\alpha$-particle with $^{93}$Nb forming a composite nucleus  $^{97}$Tc$^*$, which emits one/two neutrons to form $^{96}$Tc/$^{95}$Tc, respectively and $t$ moves in the forward direction as a spectator \cite{dm16}.

\begin{equation}
\label{eq-1}
\begin{aligned}
\text{$^7$Li $\rightarrow$ $\alpha$ + $t$} \\
\text{$\alpha$ + $^{93}$Nb  $\rightarrow$ [$^{97}$Tc$^*$]}
\text{$\rightarrow$ $^{96}$Tc + $n$,}\\
\text{E$_{th}$ = 7.3 MeV.} \\
\text{$\alpha$ + $^{93}$Nb  $\rightarrow$ [$^{97}$Tc$^*$]}
\text{$\rightarrow$ $^{95}$Tc + 2$n$,}\\
\text{E$_{th}$ = 15.6 MeV.}
\end{aligned}   
\end{equation} 

Since one or two neuron emissions are more favorable in comparison to the $p3n$-channel, one of the production routes of $^{\text{93m}}$Mo radionuclides discussed later in Eq.\ref{eq-2}, in the low energy region; it is speculated that the ICF routes play a significant role in the enhancement of Tc isotopes within 20--30 MeV energy range (Fig \ref{fig2}). Analysis of ICF and incomplete fusion fraction, (F$_{ICF}$), has been made within 20--30 MeV energy range (Fig \ref{fig3}), as is discussed later in this section. A substantial amount of ICF was observed over CF, particularly towards lower energy region, and ICF fraction is found to increase with increasing projectile energies.

The measured excitation function of $^{\text{93m}}$Mo residue populated via $\alpha$-emitting channel has been compared with various theoretical model calculations, illustrated in the Fig \ref{fig4}. The WE + HMS calculations, mostly overproduced the measured data over the entire energy range, while the HF + EM estimations underpredict them except some exception at the low energy range. Total (ground + isomeric) theoretical cross section of $^{93}$Mo has also been plotted corresponding to both EGSM and GSM as described in our previous article \cite{dm16} where in comparison with Alice91 and PACE4 was also depicted.

Scrutinizing the previous \cite{dm16} and present comparative analysis of the theoretical model calculations with the experimental results (Figs. \ref{fig1}, \ref{fig2}), it is clear that $xn$ and $pxn$ channels are best predicted by EMPIRE with EGSM level density compare to others. Moreover, the reliability of EMPIRE has also been confirmed in many other heavy ion reactions such as $^{11}$B + $^{89}$Y, $^{11}$B + $^{93}$Nb, $^{7}$Li + $^{\text{nat}}$Mo reactions \cite{ dm16,dm17,deepak17}. Therefore, it is also expected that amount of $^{\text{93m}}$Mo produced by the CF process via $\alpha$$3n$ channel in the $^{7}$Li + $^{93}$Nb reaction would be best described by the EMPIRE with EGSM similar to good reproduction of the $^{\text{93m}}$Mo and $^{97}$Ru residues populated via $\alpha$$3n$ channel in the tightly bound $^{11}$B-induced reaction on $^{89}$Y and $^{93}$Nb target, respectively.

The enhancement in the $\alpha$-emitting channel may be ascribed to ICF as several experimental investigations have now confirmed the direct breakup of weakly bound $^{7}$Li-projectile into $\alpha$ + $t$ or breakup into $\alpha$ + $\alpha$ and $\alpha$ + $d$ via different transfer processes prior to the fusion with target nucleus within $\sim$ 3--7 MeV/nucleon \cite{capurro16,pandit16,parkar16}. Thus, the following ICF processes may contribute to the large production cross section of $^{\text{93m}}$Mo in the $^{7}$Li + $^{93}$Nb reaction--

1. ICF of direct breakup component $\alpha$-particle with $^{93}$Nb forming a composite nucleus  $^{97}$Tc$^*$, which emits one proton and three neutrons to form $^{\text{93m}}$Mo, and $t$ moves in the forward direction as a spectator.
\begin{equation}
\label{eq-2}
\begin{aligned}
\text{$^7$Li $\rightarrow$ $\alpha$ + $t$} \\
\text{$\alpha$ + $^{93}$Nb  $\rightarrow$ [$^{97}$Tc$^*$]}
\text{$\rightarrow$ $^{\text{93m}}$Mo + $p$ + 3$n$,}\\
\text{E$_{th}$ = 30.75\ MeV.}
\end{aligned}   
\end{equation}     	

2. ICF of direct breakup component $t$ with $^{93}$Nb forming a composite nucleus  $^{96}$Mo$^*$, which emits three neutrons to form $^{\text{93m}}$Mo, and $\alpha$-particle moves in the forward direction as a spectator.
\begin{equation}
\label{eq-3}
\begin{aligned}
\text{$^7$Li $\rightarrow$ $\alpha$ + $t$} \\
\text{$t$ + $^{93}$Nb  $\rightarrow$ [$^{96}$Mo$^*$]} 
\text{$\rightarrow$ $^{\text{93m}}$Mo + 3$n$,}\\
\text{E$_{th}$  = 9.98 \ MeV.}
\end{aligned}  
\end{equation}  

3. Neutron stripping from the $^{7}$Li followed by ICF of breakup component $\alpha$ (from $^{6}$Li) with $^{94}$Nb forming a composite nucleus  $^{98}$Tc$^*$, which emits one proton and four neutrons to form $^{\text{93m}}$Mo, and $d$ moves in the forward direction as a spectator.
\begin{equation}
\label{eq-4}
\begin{aligned}
\text{$^7$Li + $^{93}$Nb $\rightarrow$ $^6$Li($\alpha$ + $d$) + $^{94}$Nb} \\
\text{ $\alpha$ + $^{94}$Nb $\rightarrow$ [$^{98}$Tc$^*$]} 
\text{$\rightarrow$ $^{\text{93m}}$Mo + $p$ + 4$n$,}\\
\text{E$_{th}$  = 38.28 \ MeV.} 
\end{aligned} 
\end{equation}
Moreover, ICF of $\alpha$-particle with target nucleus $^{93}$Nb may also be possible as described in equation \ref{eq-1}. 

4. Neutron stripping from the $^{7}$Li followed by ICF of breakup component $d$ (from $^{6}$Li) with $^{94}$Nb forming a composite nucleus  $^{96}$Mo$^*$, which emits three neutrons to form $^{\text{93m}}$Mo, and $\alpha$-particle moves in the forward direction as a spectator.
\begin{equation}
\label{eq-5}
\begin{aligned}
\text{$^7$Li + $^{93}$Nb $\rightarrow$ $^6$Li($\alpha$ + $d$) + $^{94}$Nb}\\
\text{$d$ + $^{94}$Nb  $\rightarrow$ [$^{96}$Mo$^*$]} 
\text{$\rightarrow$ $^{\text{93m}}$Mo + 3$n$,}\\
\text{E$_{th}$  = 10.87 \ MeV.} 
\end{aligned} 
\end{equation}
Moreover, ICF of $d$ component with target nucleus $^{93}$Nb may be possible forming a composite nucleus  $^{95}$Mo$^*$, which emits two neutrons to form $^{\text{93m}}$Mo, and $\alpha$-particle moves in the forward direction as a spectator. 
\begin{equation} 
\label{eq-6}        
\begin{aligned}
\text{$d$ + $^{93}$Nb  $\rightarrow$ [$^{95}$Mo$^*$]} 
\text{$\rightarrow$ $^{\text{93m}}$Mo + 2$n$,}\\
\text{E$_{th}$  = 3.49 \ MeV.} 
\end{aligned} 
\end{equation}            

5. Proton pickup by the $^{7}$Li followed by ICF of breakup component $\alpha$ (from $^{8}$Be) with $^{92}$Zr forming a composite nucleus  $^{96}$Mo$^*$, which emits three neutrons to form $^{\text{93m}}$Mo, and another $\alpha$-particle moves in the forward direction as a spectator.
\begin{equation}
\label{eq-7}
\begin{aligned}
\text{$^7$Li + $^{93}$Nb $\rightarrow$ $^8$Be($\alpha$ + $\alpha$) + $^{92}$Zr}\\ 
\text{$\alpha$ + $^{92}$Zr  $\rightarrow$ [$^{96}$Mo$^*$]} 
\text{$\rightarrow$ $^{\text{93m}}$Mo + 3$n$,}\\
\text{E$_{th}$  = 24.46 \ MeV.} 
\end{aligned} 
\end{equation} 
Moreover, ICF of $\alpha$-particle with target nucleus $^{93}$Nb may also be possible as described in equation \ref{eq-1}.

Hence, the enhancement in the measured cross section over the EMPIRE estimation, which does not consider the ICF process, may be regarded as ICF as long as it explains the other reaction channels with the same set of parameters. The excitation functions of CF--ICF of $^{\text{93m}}$Mo and the fraction of incomplete fusion (F$_{ICF}$) at different projectile energies have been represented in Figs. \ref{fig5}(a) and (b), respectively. The ICF for $^{\text{93m}}$Mo and Tc isotopes (i.e., sum of $^{96}$Tc and $^{95}$Tc) was estimated from the difference of their experimental and theoretical cross sections using
	 $\sigma^{^\text{{93m}Mo}}_{ICF} = \sigma^{^\text{{93m}Mo}}_{expt} - \sigma^{^\text{{93m}Mo}}_{theor}$, 
	 and
	 	 $\sigma^{^\text{Tc}}_{ICF} = \sigma^{^\text{Tc}}_{expt} - \sigma^{^\text{Tc}}_{theor}$ expressions, respectively, 
	  where $\sigma_{expt}$ and $\sigma_{theor}$ represent experimental and theoretical cross section for $^{\text{93m}}$Mo and Tc isotopes.
The ICF fraction (F$_ {ICF} $) is a measure of the extent of ICF relative to total reaction cross section and defined as F$_{\text{ICF}} = (\sigma_{ICF}/\sigma_{react}^{theor})\times$100, where $\sigma_{react}^{theor}$ is the total theoretical reaction cross section which is equivalent to theoretical fusion cross section in this case \cite{dm17}. The increasing trend of ICF fraction of $^{\text{93m}}$Mo with increasing projectile energies was observed within 4.5--6.5 MeV/nucleon energy in the $^{7}$Li + $^{93}$Nb, similar to that observed in other weakly bound and $\alpha$-cluster projectiles \cite{dm17,abhishek12,sharma14}. However, the short half-lives of the residual radionuclides, or production of stable isotopes could not be identified with this method; hence the ICF corresponding to these missing channels can not be determined. Thus, the computed ICF cross section (Fig. \ref{fig5}(a)) may be considered as the lower limit of ICF for the $^{7}$Li + $^{93}$Nb reaction in the $\alpha$-emitting channels.

   	\section{\label{s4}Summary}
		
		This article describes the analysis of ICF in the $^{7}$Li-induced reaction on $^{93}$Nb target over CF within 20--45 MeV energy range. A sum of the cross section measured in the neutron and proton emitting channels at the various projectile energies is fairly reproduced by the EMPIRE3.2 with EGSM in comparison to the Alice14. Perhaps simplified coupled channel calculation together with HF and EM for EQ and PEQ processes including collective effect dependent EGSM/GSM level density make the EMPIRE reliable for heavy-ion induced reactions.
		An enhancement in the measured excitation function of the proton (20--30 MeV) and $\alpha$-emitting channels was observed compare to the EMPIRE estimation with the EGSM level density using the same set of parameters, which excellently reproduce the $xn$, and $pxn$-channels towards higher energy region within experimental uncertainties. The enhancement in the $^{95,96}$Tc and $^{\text{93m}}$Mo production may be attributed to ICF due to the projectile ($^{7}$Li) breakup into $\alpha$-particle + $t$, or $\alpha$ + $\alpha$ and $\alpha$-particle + $d$ following transfer processes between the interacting nuclei.




	\begin{acknowledgments}
		The authors convey their sincere thanks to S. Lahiri, SINP, Kolkata, and the Pelletron staff of the BARC-TIFR Pelletron facility, Mumbai, for their cooperation and help during the main experiment. DK gratefully acknowledges the financial support of MHRD, Government of India.
	\end{acknowledgments}

\end{document}